\documentclass[aps,showpacs]{revtex4}
\usepackage[utf8]{inputenc}
\usepackage{color}
\usepackage{amsmath}
\usepackage{graphicx}

\makeatletter
\@ifundefined{textcolor}{}
{%
 \definecolor{BLACK}{gray}{0}
 \definecolor{WHITE}{gray}{1}
 \definecolor{RED}{rgb}{1,0,0}
 \definecolor{GREEN}{rgb}{0,1,0}
 \definecolor{BLUE}{rgb}{0,0,1}
 \definecolor{CYAN}{cmyk}{1,0,0,0}
 \definecolor{MAGENTA}{cmyk}{0,1,0,0}
 \definecolor{YELLOW}{cmyk}{0,0,1,0}
}

\usepackage{dcolumn}
\usepackage{bm}
\usepackage{wrapfig}
\usepackage{subfigure}
\usepackage{ucs}
\usepackage{lineno}
\usepackage{epsfig}
\usepackage{psfrag}\usepackage{epstopdf}
\DeclareGraphicsExtensions{.pdf,.eps,.png,.jpg,.mps}

\makeatother

\begin{document}
\title{Impurity-induced increase in the thermal quantum correlations and teleportation in an Ising-$XXZ$ diamond chain }
\author{Saulo L. L. Silva$^{1}$, Moises Rojas$^{2}$}
\affiliation{$^{1}$Centro Federal de Educação Tecnológica de Minas Gerais , 37250-000,
Nepomuceno-MG, Brazil}
\affiliation{$^{2}$Departamento de Física, Universidade Federal de Lavras, 37200-900,
Lavras-MG, Brazil}
\begin{abstract}
In this work we analyze the quantum correlations in a spin-$1/2$ Ising-$XXZ$ diamond chain with one plaquette distorted impurity. We have shown that the introduction of impurity into the chain can significantly increase entanglement as well as quantum correlations compared to the original model, without impurity. Due to the great flexibility in the choice of impurity parameters, the model presented is very general and this fact can be very useful for future experimental measurements. In addition to entanglement and quantum coherence, we studied quantum teleportation through a quantum channel composed by a coupled of Heisenberg dimers with distorted impurity in an Ising-$XXZ$ diamond chain, as well as  fidelity in teleportation. Our analysis shows that the appropriate choice of parameters can greatly increase all the measures analyzed. For comparison purposes, we present all our results together with the results of the measurements made for the original model, without impurity,  studied in previous works.
\end{abstract}


\maketitle

\section{Introduction}
\label{intro}
Quantum correlations not only has interesting properties of quantum
mechanics but also is very important due to the powerful applications
in quantum information process and quantum computing \cite{Ben1,Ben2,lamico,cirac}.

In the last decades, the spin chains with Heisenberg interaction have
received attention due to the fact that they are promising candidates
for physical implementation of quantum information processing \cite{loss,apo-1,ame,divi,bose,kam,zhou}.

On the other hand, quantum coherence that originates from the superposition
principles states is one of the central concepts in quantum mechanical
systems \cite{levi,stre,winter}. Recently various approaches have
been put forward to develop a resource theory of coherence on Heisenberg
spin models \cite{kar,wu}. Several measures of coherence have been
proposed, and their properties have been investigated in detail \cite{baum,Hu,xi,tan}.

Recently, the spin-1/2 Heisenberg diamond chain, as well as an exactly
solvable Ising-Heisenberg have been largely explored. The motivation
to research the Ising-Heisenberg diamond chain model is based in fact
that over-simplification, the generalization version of the spin-1/2
ising-Heisenberg diamond chain qualitatively reproduces thermodynamics
data reported on the real material $\mathrm{Cu_{3}\left(CO_{3}\right)_{2}\left(OH\right)_{2}}$
known as \textit{azurite}. Owing to this fact, a lot of attention
has been paid to a rigorous treatment of various versions of the Ising-Heisenberg
diamond chain \cite{kiku,cano,boh,val}. Furthermore, the thermal
entanglement of the Ising-Heisenberg diamond chain was studied in
\cite{moi,moi2,moi3,moi4,gao,cheng}. Rojas \textit{et. al. }used
the standard teleportation protocol to study an arbitrary entangled
state teleportation through a couple of Heisenberg dimers in an infinite
Ising-$XXZ$ diamond chain \cite{tele}.

The impurity plays an important role in solid state physics \cite{fal,apo}.
The impurities can be understood by a local bound disorder of the
interchain exchange coupling and this changes in the structure affects
strongly the quantum correlations. The impurity effects on quantum
entanglement have been considered in Heisenberg spin chain \cite{sale,osen,solano,sun,fub}.
Recently, the quantum discord \cite{gon} and dissipative effects
\cite{ming} based in Heisenberg chain with impurities has also been
studied. More recently,  the research in this field was extended to
the undestanding of thermal quantum entanglement \cite{mr} and quantum
correlations in Ising-Heisenberg diamond chain with impurities \cite{mro}.
Besides spin impurity is possible to include a variety of spin chains
with magnetic impurities \cite{fu,gal,brene}.

In this sense, the main goal of this work is the investigate the thermal
entanglement and quantum coherence on exactly solvable spin-1/2
Ising-$XXZ$ diamond chain with one distorted impurity plaquette inserted in
the structure. We will examine in detail the thermal entanglement
and quantum coherence, which exhibits a clear performance improvement
when manipulate the impurity compared to the original model without
impurity. On the other hand, our results show the impurity are helpful
not only for improving the quality of teleportation, but also for
enhancing the critical temperature for valid teleportation.

This paper is organized as follows. In Section II we describe the
physical model and the method of its exact treatment. In Section III,
is given a brief review concerning the definition of concurrence $\mathcal{C}$
as well as the quantum coherence quantifier $l_{1}$-norm, and next
analytical expression is found for the concurrence and $l_{1}$-norm,
respectively. We analyzed the effects of impurity parameters and several
parameter of the model on thermal entanglement and $l_{1}$-norm.
In Section IV, we briefly review the protocol of the quantum teleportation.
Then the average fidelity behavior our distorted diamond chain model
with impurity and of the original model without are investigated detailed.
Finally, in Section V, we summarize our conclusions. 
\section{The model and Approach}
\label{sec:1}
\begin{figure}\centering
	
	\includegraphics[scale=0.50]{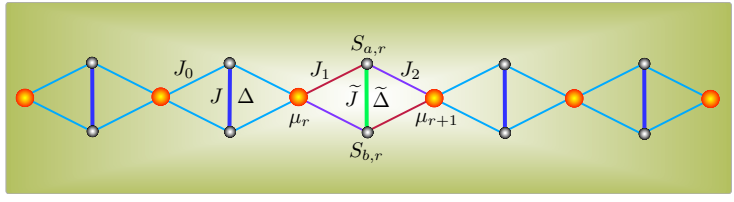}\caption{\label{fig:1}Schematic representation of the spin-1/2 Ising-$XXZ$ diamond chain with distorted impurity. The Ising spins are denoted by $\mu_{i}$ and the Heisenberg spins are represented by $S_{a(b),i}$ }
	
\end{figure}

Our model consists of a infinite spin-1/2 Ising-$XXZ$ diamond
chain with one distorted impurity plaquette under an external magnetic field
$B$, which is schematically depicted in Fig.\ref{fig:1}. The total
hamiltonian is given by

\begin{equation}
\mathcal{H}=\sum_{i=1}^{N}\mathcal{H}_{i},\label{eq:1}
\end{equation}
where 
$$\mathcal{H}_{i}=\mathcal{H}_{i}^{host}+\mathcal{H}_{i}^{imp},$$
whith

\[
\begin{array}{cl}
\mathcal{H}_{i}^{host}= & J\left(\mathbf{S}_{a,i},\mathbf{S}_{b,i}\right)_{\Delta}+J_{0}\left(S_{a,i}^{z}+S_{b,i}^{z}\right)\left(\mu_{i}+\mu_{i+1}\right)\\
& -B\left(S_{a,i}^{z}+S_{b,i}^{z}\right)-\frac{B}{2}\left(\mu_{i}+\mu_{i+1}\right),\\
& \mathrm{for}\:i=1,2,\ldots,r-1,r+1,\ldots,N
\end{array}
\]
where $\left(\mathbf{S}_{a,i},\mathbf{S}_{b,i}\right)_{\Delta}=S_{a,i}^{x}S_{b,i}^{x}+S_{a,i}^{y}S_{b,i}^{y}+\Delta S_{a,i}^{z}S_{b,i}^{z}$, corresponds to the interstitial anisotropic Heisenberg spins coupling ($J$ and $\Delta$), while the nodal-interstitial (dimer-monomer) spins $\mu_{i}=\pm1/2$ are representing the Ising-type exchanges ($J_{0}$).  \\
And

\[
\begin{array}{cl}
\mathcal{H}_{i}^{imp}= & \widetilde{J}\left(\mathbf{S}_{a,i},\mathbf{S}_{b,i}\right)_{\widetilde{\Delta}}+\left(J_{1}\mu_{i}+J_{2}\mu_{i+1}\right)S_{a,i}^{z}
+\left(J_{1}\mu_{i+1}+J_{2}\mu_{i}\right)S_{b,i}^{z}\\
&-B\left(S_{a,i}^{z}+S_{a,i}^{z}\right)
-\frac{B}{2}\left(\mu_{i}+\mu_{i+1}\right),\\
& \mathrm{for}\:i=r,
\end{array}
\]
is the Hamiltonian of the impurity. Here $J_{1}=J_{0}(1+\eta)$, $J_{2}=J_{0}(1.0+\gamma)$, $\widetilde{J}=J(1.0+\alpha)$,
$\widetilde{\Delta}=\Delta(1.0+\Omega)$ represents the impurity parameters. Throughout 
the text, the symbol \, $\widetilde{}$ \, on the parameter indicates that this parameter is related to 
impurity in the chain.

The eigenvalues of $\mathcal{H}_{i}^{host}$ is given by

\[
\begin{array}{cl}
\varepsilon_{1,4}= & \frac{J\Delta}{4}\pm\left(J_{0}\mp\frac{B}{2}\right)\left(\mu_{i}+\mu_{i+1}\right)\mp\frac{B}{2},\\
\varepsilon_{2,3}= & -\frac{J\Delta}{4}\pm\frac{J}{2}-\frac{B}{2}\left(\mu_{i}+\mu_{i+1}\right),
\end{array}
\]
and the corresponding eigenvectors are
\begin{eqnarray}
|\varphi_{i1}\rangle & = & |00\rangle_{i},\\
|\varphi_{i2,i3}\rangle & = & \frac{1}{\sqrt{2}}\left(|01\rangle_{i}\pm|10\rangle_{i}\right),\\
|\varphi_{i1}\rangle & = & |11\rangle_{i}.
\end{eqnarray}

Likewise eigenvalues of $\mathcal{H}_{i}^{imp}$ is given by

\[
\begin{array}{cl}
\widetilde{\varepsilon}_{1,4}= & \frac{\widetilde{J}\widetilde{\Delta}}{4} \pm \left(\frac{(J_{1}+J_{2})}{2}\mp\frac{B}{2}\right)\left(\mu_{r}+\mu_{r+1}\right)\mp B,\\
\widetilde{\varepsilon}_{2,3}= & -\frac{\widetilde{J}\widetilde{\Delta}}{4}-\frac{B}{2}\left(\mu_{r}+\mu_{r+1}\right)\pm\frac{1}{2}\sqrt{\Sigma^{2}+\widetilde{J}^{2}},
\end{array}
\]
where $\Sigma=\left(J_{1}-J_{2}\right)\left(\mu_{r}-\mu_{r+1}\right)$.
The corresponding eigenvectors are given by

\begin{eqnarray}
|\widetilde{\varphi}_{r1}\rangle & =&|00\rangle_{r},\\
|\widetilde{\varphi}_{r2,r3}\rangle & =&M_{\pm}|01\rangle_{r}+N_{\pm}|10\rangle_{r},\\
|\widetilde{\varphi}_{r4}\rangle & =&|11\rangle_{r},
\end{eqnarray}
where 
$$M_{\pm}=\frac{\widetilde{J}}{\sqrt{2\widetilde{J}^2+2\Sigma^2 \mp2\Sigma \sqrt{\Sigma^2 +\widetilde{J}^2}}},$$
and 
$$N_{\pm}=\frac{-\Sigma \pm \sqrt{\Sigma^2 + \widetilde{J}^2}}{\sqrt{2\widetilde{J}^2+2\Sigma^2 \mp2\Sigma \sqrt{\Sigma^2 +\widetilde{J}^2}}}.$$  
Notice that, in general $J_1 \neq J_2$, therefore $\mathcal{H}_{i}^{imp}$ has diferent 
eigenvector from $\mathcal{H}_{i}^{host}.$ In addition, the probabilities of obtaining 
$|01\rangle_{r}$ and $|10\rangle_{r}$ are different in eigenstate $|\widetilde{\varphi}_{r2,r3}\rangle$. 
This is the main difference between our results and those presented in previous works \cite{moi,mr,mro}.

\subsection{The partition function}
\label{sec:2}
The state at thermal equilibrium can be described by the Gibb's density
operators $\rho(T)=\frac{\exp(-\beta \mathcal{H})}{Z}$, where $\beta=1/k_{B}T$,
with $k_{B}$ being the Boltzmann's constant, $T$ is the absolute
temperature, whereas the partition function of the system is defined
by $Z=Tr\left[\exp(-\beta \mathcal{H})\right]$. Similarly to that made by 
Freitas $\mathit{et\,al}$ \cite{mro} we consider the chain with periodic boundary 
condition and use the transfer-matrix notation to obtain the partition function. 
Thus

\[
\begin{array}{cl}
Z = & \sum_{\{\mu\}}w(\mu_1,\mu_2)...w(\mu_{r-1},\mu{r})\widetilde{w}(\mu_{r},\mu_{r+1})
w(\mu_{r+1},\mu_{r+2})...w(\mu_{N},\mu_{1}),
\end{array}
\]
where 

\begin{equation}
w(\mu_i,\mu_{i+1})=\sum_{j=1}^{4}e^{-\beta \epsilon_{ij}(\mu_i,\mu_{i+1})},
\end{equation}
and
\begin{equation}
\widetilde{w}(\mu_i,\mu_{i+1})=\sum_{j=1}^{4}e^{-\beta \widetilde{\epsilon}_{ij}(\mu_i,\mu_{i+1})}.
\end{equation}
The partition function can be simplified by taking $Z=\mathrm{Tr}(\widetilde{W}W^{N-1})$, 
in such a way that
\[
W=\left[\begin{array}{cc}
w(\frac{1}{2},\frac{1}{2}) & w(\frac{1}{2},-\frac{1}{2})\\
w(-\frac{1}{2},\frac{1}{2}) & w(-\frac{1}{2},-\frac{1}{2})
\end{array}\right],
\]
and
\[
\widetilde{W}=\left[\begin{array}{cc}
\widetilde{w}(\frac{1}{2},\frac{1}{2}) & \widetilde{w}(\frac{1}{2},-\frac{1}{2})\\
\widetilde{w}(-\frac{1}{2},\frac{1}{2}) & \widetilde{w}(-\frac{1}{2},-\frac{1}{2})
\end{array}\right].
\]
In order to simplify the notation we will start to denote $w_{\pm \pm} \equiv w(\pm \frac{1}{2},\pm \frac{1}{2})$ 
and $w_{\pm \mp} \equiv w(\pm \frac{1}{2},\mp \frac{1}{2})$. Through the 
diagonalization of the transfer-matrix we obtain

\begin{equation}
Z = a\Lambda_{+}^{N-1}+d\Lambda_{-}^{N-1},
\end{equation}
where $$\Lambda_{\pm} =\frac{w_{++}+w_{--}\pm Q}{2}$$ with $Q =\sqrt{(w_{++}-w_{--})^2+4w_{+-}^2}$
are the eigenvalues associated with the matrix W. Furthermore

\begin{equation}
a =\frac{4w_{+-}\widetilde{w}_{+-}+(w_{++}-w{--})(\widetilde{w}_{++}-\widetilde{w}_{--})+Q(\widetilde{w}_{++}+\widetilde{w}_{--})}{2Q},
\end{equation}
and
\begin{equation}
d = \frac{-4w_{+-}\widetilde{w}_{+-}-(w_{++}-w{--})(\widetilde{w}_{++}-\widetilde{w}_{--})+Q(\widetilde{w}_{++}+\widetilde{w}_{--})}{2Q}.
\end{equation}


At the thermodynamic limit, $N\rightarrow \infty$, we have $$Z=a\Lambda_{+}^{N-1},$$
since $a\Lambda_{+}^{N-1} > d\Lambda_{-}^{N-1}.$
\subsection{Average reduced density operator}
\label{sec:3}
In order to obtain the reduced density matrix let's take the
thermal average for each two-qubit Heisenberg\cite{mr,mro},.
We start by defining the operator as a function of spin particles $\mu_{i}$ and $\mu_{i+1}$

\begin{equation}
\varrho(\mu_{i},\mu_{i+1})=\sum_{i=1}^{4} e^{-\beta \epsilon_{i,j}}|\phi_{i,j}\rangle \langle \phi_{i,j}|.
\end{equation}

For impurity we define the operator $\widetilde{\varrho}$ of the two-qubit Heisenberg operator

\[
\widetilde{\varrho}(\mu_{r},\mu_{r+1})=\left[\begin{array}{cccc}
\widetilde{\varrho}_{1,1} & 0 & 0 & 0\\
0 & \widetilde{\varrho}_{2,2} & \widetilde{\varrho}_{2,3} & 0\\
0 & \widetilde{\varrho}_{3,2} & \widetilde{\varrho}_{3,3} & 0\\
0 & 0 & 0 & \widetilde{\varrho}_{4,4}
\end{array}\right],
\]
where

\[
\begin{array}{cl}
\widetilde{\varrho}_{1,1}(\mu_{r},\mu_{r+1})= & \mathrm{e}^{-\beta\widetilde{\varepsilon}_{r1}},\\
\widetilde{\varrho}_{2,2}(\mu_{r},\mu_{r+1})= & \mathrm{e}^{-\beta\widetilde{\varepsilon}_{r2}}M_{+}^{2}+\mathrm{e}^{-\beta\widetilde{\varepsilon}_{r3}}M_{-}^{2},\\
\widetilde{\varrho}_{2,3}(\mu_{r},\mu_{r+1})= & \mathrm{e}^{-\beta\widetilde{\varepsilon}_{r2}}M_{+}N_{+}+\mathrm{e}^{-\beta\widetilde{\varepsilon}_{r3}}M_{-}N_{-},\\
\widetilde{\varrho}_{4,4}(\mu_{r},\mu_{r+1})= & \mathrm{e}^{-\beta\widetilde{\varepsilon}_{r4}}.
\end{array}
\]

Using the transfer-matrix approach we can write the elements 
of the density matrix for impurity $\widetilde{\rho}_{k,l}$ in the form \cite{mro}

\begin{equation}\label{eqrho}
\widetilde{\rho}_{k,l}=\frac{1}{Z}\mathrm{Tr}\left(\widetilde{P}_{k,l}W^{N-1} \right),
\end{equation}
where

\[
\widetilde{P}_{k,l}=\left[\begin{array}{cc}
\widetilde{\rho}_{k,l}(\frac{1}{2},\frac{1}{2}) & \widetilde{\rho}_{k,l}(\frac{1}{2},-\frac{1}{2})\\
\widetilde{\rho}_{k,l}(-\frac{1}{2},\frac{1}{2}) & \widetilde{\rho}_{k,l}(-\frac{1}{2},-\frac{1}{2})
\end{array}\right],
\]
consider $\widetilde{\rho}_{k,l}(\pm \pm) \equiv \widetilde{\rho}_{k,l}(\pm \frac{1}{2},\pm \frac{1}{2})$ and $\widetilde{\rho}_{k,l} (\pm \mp) \equiv \widetilde{\rho}_{k,l}(\pm \frac{1}{2},\mp \frac{1}{2})$.
Being $U$ the matrix that diagonalizes $W$ we can rewrite Eq. (\ref{eqrho}) in the form

\[
\begin{array}{ccc}
\widetilde{\rho}_{k,l} & = & \frac{\mathrm{Tr}\left(U^{-1}\widetilde{P}_{k,l}U 
	\left[\begin{array}{cc}
	\Lambda_{+}^{N-1} & 0\\
	0 & \Lambda_{-}^{N-1}
	\end{array}\right]
	\right)}{a\Lambda_{+}^{N-1}+d\Lambda_{-}^{N-1}},\end{array}
\]
that at the thermodynamic limit, $N \rightarrow \infty$, it reduces to
\[
\begin{array}{ccc}
\widetilde{\rho}_{k,l} & = & \frac{\mathcal{A}_{k,l}+\mathcal{B}_{k,l}}{\mathcal{M}_{k,l}},\end{array}
\]
where 

\[
\begin{array}{cl}
\mathcal{A}_{k,l}= & Q\left[\widetilde{\varrho}_{k,l}(++)+\widetilde{\varrho}_{k,l}(--)\right]+4w_{+-}\widetilde{\varrho}_{k,l}(+-)\\
\mathcal{B}_{k,l}= & \left[\widetilde{\varrho}_{k,l}(++)-\widetilde{\varrho}_{k,l}(--)\right]\left(w_{++}-w_{--}\right)\\
\mathcal{M}= & Q\left(\widetilde{w}_{++}+\widetilde{w}_{--}\right)+4w_{+-}\widetilde{w}_{+-}+
\left(\widetilde{w}_{++}-\widetilde{w}_{--}\right)\left(w_{++}-w_{--}\right).
\end{array}
\]

Finally, we conclude that the elements of the reduced density matrix take the form

\begin{equation}
\widetilde{\rho}(T)=\left[\begin{array}{cccc}
\widetilde{\rho}_{11} & 0 & 0 & 0\\
0 & \widetilde{\rho}_{22} & \widetilde{\rho}_{23} & 0\\
0 & \widetilde{\rho}_{23} & \widetilde{\rho}_{33} & 0\\
0 & 0 & 0 & \widetilde{\rho}_{44}
\end{array}\right].\label{eq:5}
\end{equation}

The state at thermal equilibrium can be described by the Gibb's density
operators $\rho(T)$ represents a thermal density operator, the entanglement
in the thermal state is simply called thermal entanglement.

\section{Quantum Correlations}
\label{sec:4}
In order to describe the thermal entanglement of
two-qubit system, the concurrence is used as a measure of the entanglement.
The concurrence $\mathcal{C}$ is defined as \cite{wootters} 
\begin{eqnarray*}
	\mathcal{C}={\rm {max}\left\{ 0,\sqrt{\lambda_{1}}-\sqrt{\lambda_{2}}-\sqrt{\lambda_{3}}-\sqrt{\lambda_{4}}\right\} ,}
\end{eqnarray*}
where $\lambda_{i}\:(i=1,2,3,4)$ are the eigenvalues in decreasing
order of the matrix 
\begin{eqnarray*}
	R=\rho\left(\sigma^{y}\otimes\sigma^{y}\right)\rho^{\ast}\left(\sigma^{y}\otimes\sigma^{y}\right),
\end{eqnarray*}
with $\sigma^{y}$ being the Pauli matrix. We will explore the thermal
entanglement of the model. To this end, we should obtain the eigenvalues
of $R$. Thus, the concurrence $\mathcal{C}$ can be written as

\[
\mathcal{C}(\widetilde{\rho})=2\mathrm{max}\{|\widetilde{\rho}_{2,3}|-\sqrt{\widetilde{\rho}_{1,1}\widetilde{\rho}_{4,4}},0\}\;.
\]
In this case, the analytical expression of the thermal concurrence
is too large to be explicitly provided in this paper.

In addition to entanglement, the system may also have other quantum
correlations, such as quantum discord. Quantum coherence is more robust
than entanglement, so it can be a key ingredient in the development
of quantum information. We calculate the quantum coherence by the
$l_{1}$-norm defined by the sum of the absolute values of the elements
off the main diagonal of the reduced density operator of the system
\cite{baum}. So 
\[
\mathcal{C}_{l_{1}}(\rho)=\sum_{i\neq j}|\langle i|\rho|j\rangle|.
\]

\subsection{Concurrence}
\label{sec:5}
In our model $J_{1}$ and $J_{2}$ they can be taken independently.
This guarantees a much greater generality of this model in relation
to those previously studied \cite{moi,mr,mro}. This greater generality
is of great interest in experimental measurements since it allows
greater flexibility in modeling.
\begin{figure}\centering
	\includegraphics[scale=0.38]{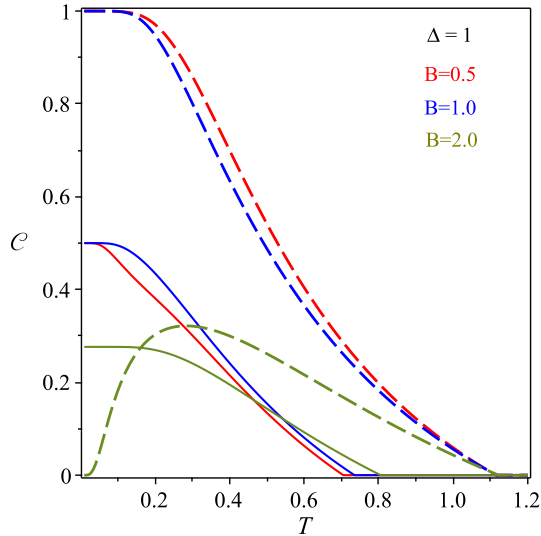} \caption{\label{fig:2}The concurrence $\mathcal{C}$ as a function of temperature
		$T$ for $J_{0}/J=1.0$, $\Delta=1.0$, $\alpha=0.0$, solid curve
		($\eta=0.0$, $\gamma=0.0$, $\Omega=0.0$), dashed curve ($\eta=-0.5$,
		$\gamma=-0.6$, $\Omega=0.8$).}
\end{figure}

With an appropriate choice of parameters, we can obtain optimal entanglement
results. In Fig. \ref{fig:2}, we show the concurrence $\mathcal{C}$
as a function of temperature $T$ and for different values of magnetic
field. The solid curve indicates the original model (without impurity)
while the dashed curve indicates the model with impurity described
by the Hamiltonian $\mathcal{H}$. In addition, the red curve refers
to a magnetic field $B=1.0$ and the blue curve refers to $B=2.0$.
For the impurity model, the adopted parameters were ($\eta=-0.5$,
$\gamma=-0.6$, $\Omega=0.8$). In relation to the original model
\cite{moi}, we can clearly see the advantage of the model with impurity.
The impurity model presents greater entanglement and also a greater
critical entanglement temperature. For higher fields we can observe
a different behavior for the two models. In this case, the entanglement
in the model with impurity is zero for $T=0.0$ and different from
zero for higher temperatures, until it becomes null at the critical
entanglement temperature. In this case, the triplet splits and $|\widetilde{\varphi}_{r_{1},r_{2}}\rangle$
becames the ground states. In this case there is no entanglement at
$T=0.0$. But the increase in temperature increases entanglement by
bringing in some singlet component into the mixture.
\begin{figure}\centering
	\includegraphics[scale=0.38]{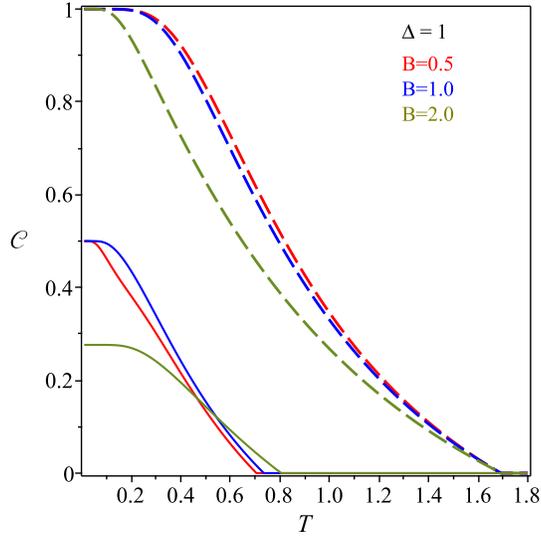}
	
	\caption{\label{fig:3}The concurrence $\mathcal{C}$ as a function of the
		temperature $T$, $J_{0}/J=1.0$ , $\Delta=1.0$, solid curve(original
		model), dashed curve ($\alpha=0.5$, $\eta=-0.5$, $\gamma=-0.6$,
		$\Omega=0.8$). }
\end{figure}

\begin{figure}\centering
	\includegraphics[scale=0.50]{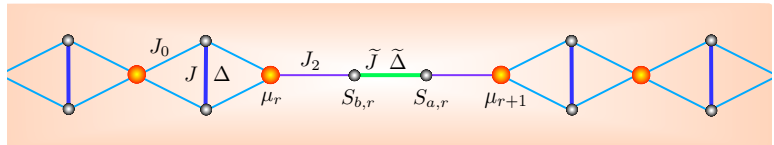} \caption{\label{fig:4}Schematic representation of the form taken by the impurity
		when we take $\eta=-1.0$ ($J_{1}=0.0$). Note that instead of the
		diamond shape, the impurity structure now takes on a horizontal shape.}
\end{figure}

In Figure \ref{fig:3} we analyze how a change in the interaction
between the spins of the impurity affects the entanglement. This interaction
is represented by $\widetilde{J}$ and was obtained by taking $\alpha=0.5$.
In other words, the interaction between the Heisenberg spins in the
impurity was greater than between the Heisenberg spins in the rest
of the chain. The advantage of this small modification is evident
in relation to the situation analyzed in Figure \ref{fig:2}. In this
case we can see that the entanglement is different from zero even
for high field. In addition, the increase in the critical entanglement
temperature was very significant compared to the original model (solid
curve).

\begin{figure}
	\includegraphics[scale=0.30]{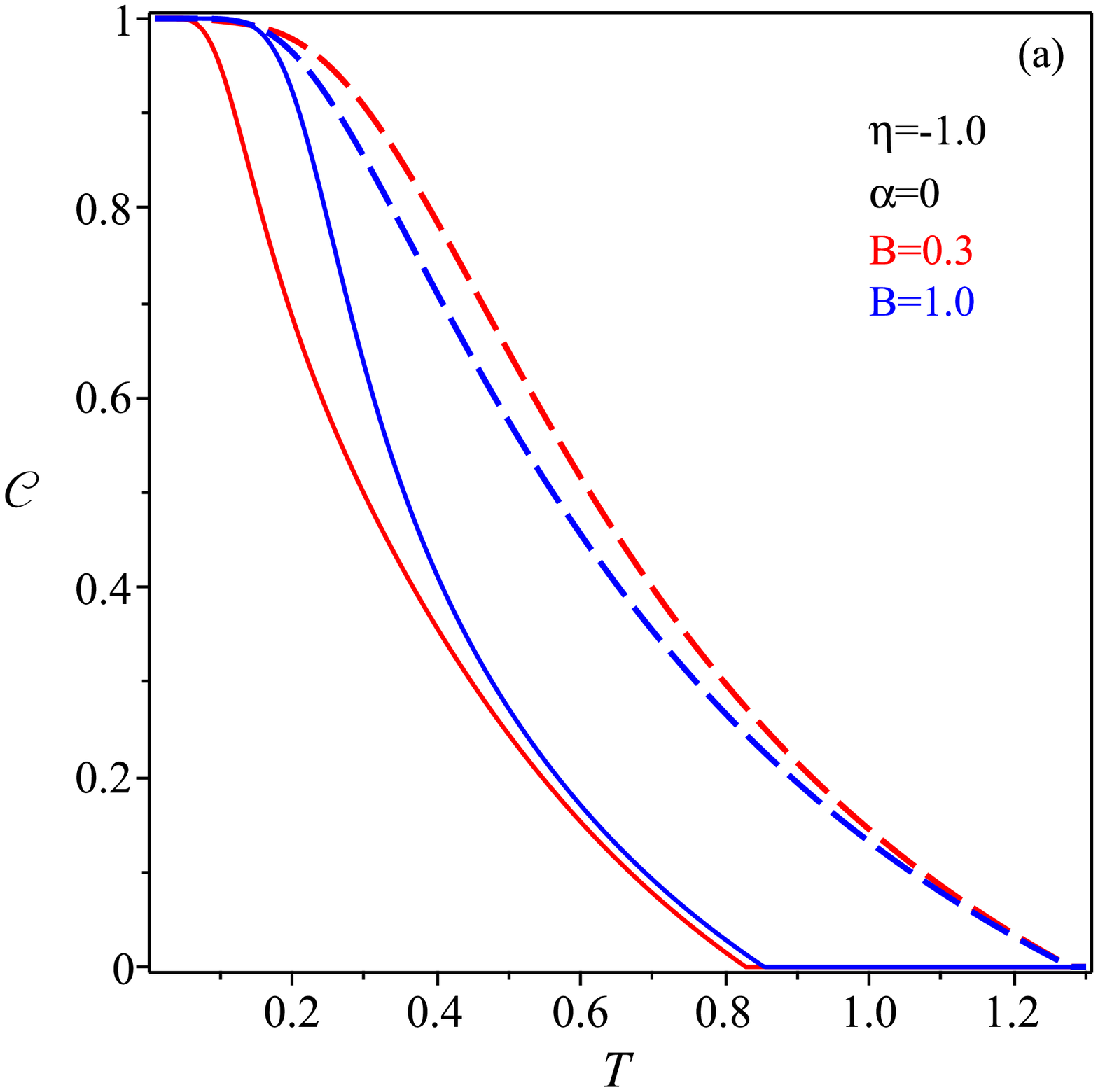}\includegraphics[scale=0.30]{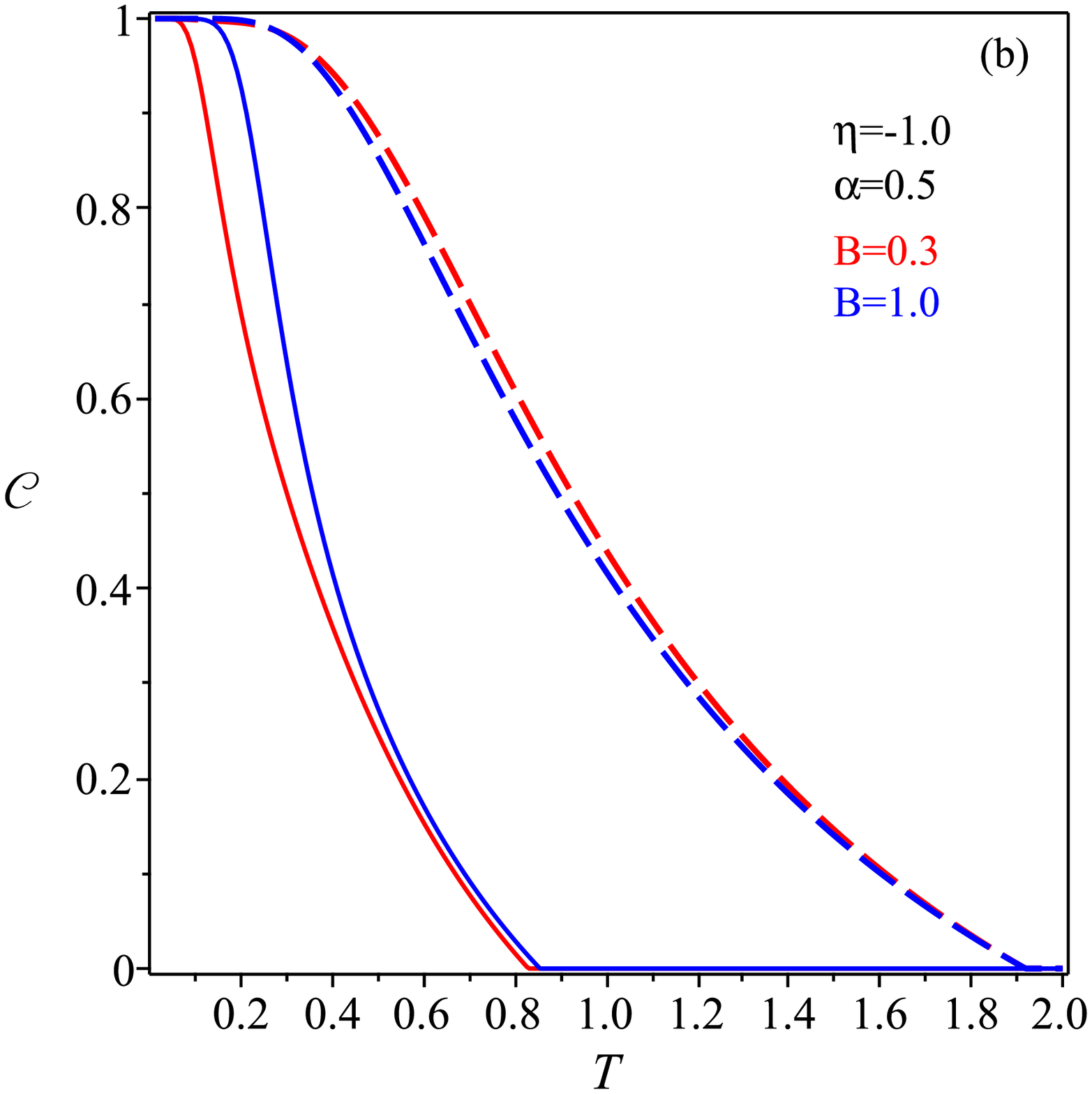}
	
	\caption{\label{fig5.1}The concurrence $\mathcal{C}$ as a function of the
		temperature $T$, we set $J_{0}/J=1.0$, $\Delta=1.3$, $\eta=-1.0$,
		(a) dashed curve ($\alpha=0.0$, $\gamma=-0.8$, $\Omega=0.8$), (b)
		dashed curve ($\alpha=0.5$, $\gamma=-0.8$, $\Omega=0.8$)}
\end{figure}

We observe an interesting behavior in the system when we take $J_{1}=0.0$
or $\eta=-1.0$. This condition is equivalent to a change in the structure
of the impurity. Instead of the diamond shape shown in Figure \ref{fig:1},
the impurity term have a horizontal interaction as shown in Figure
\ref{fig:4}. This situation is analyzed in Figure \ref{fig5.1} with
the following parameters $J_{0}/J=1.0$, $\Delta=1.3$, $\eta=-1.0$.
In Figure \ref{fig5.1}(a) we considered $\alpha=0.0$ ($J=\widetilde{J}$)
and we observed entanglement and critical entanglement temperature
higher than the original model (solid curve). In Figure \ref{fig:3}
we see an increase in entanglement with $\alpha=0.5$, the same we
observed in this case. In Figure \ref{fig5.1} (b) is possible observe
the entanglement gain in relation to the original model (solid curve),
both for low and high fields.


\begin{figure}
	\includegraphics[scale=0.30]{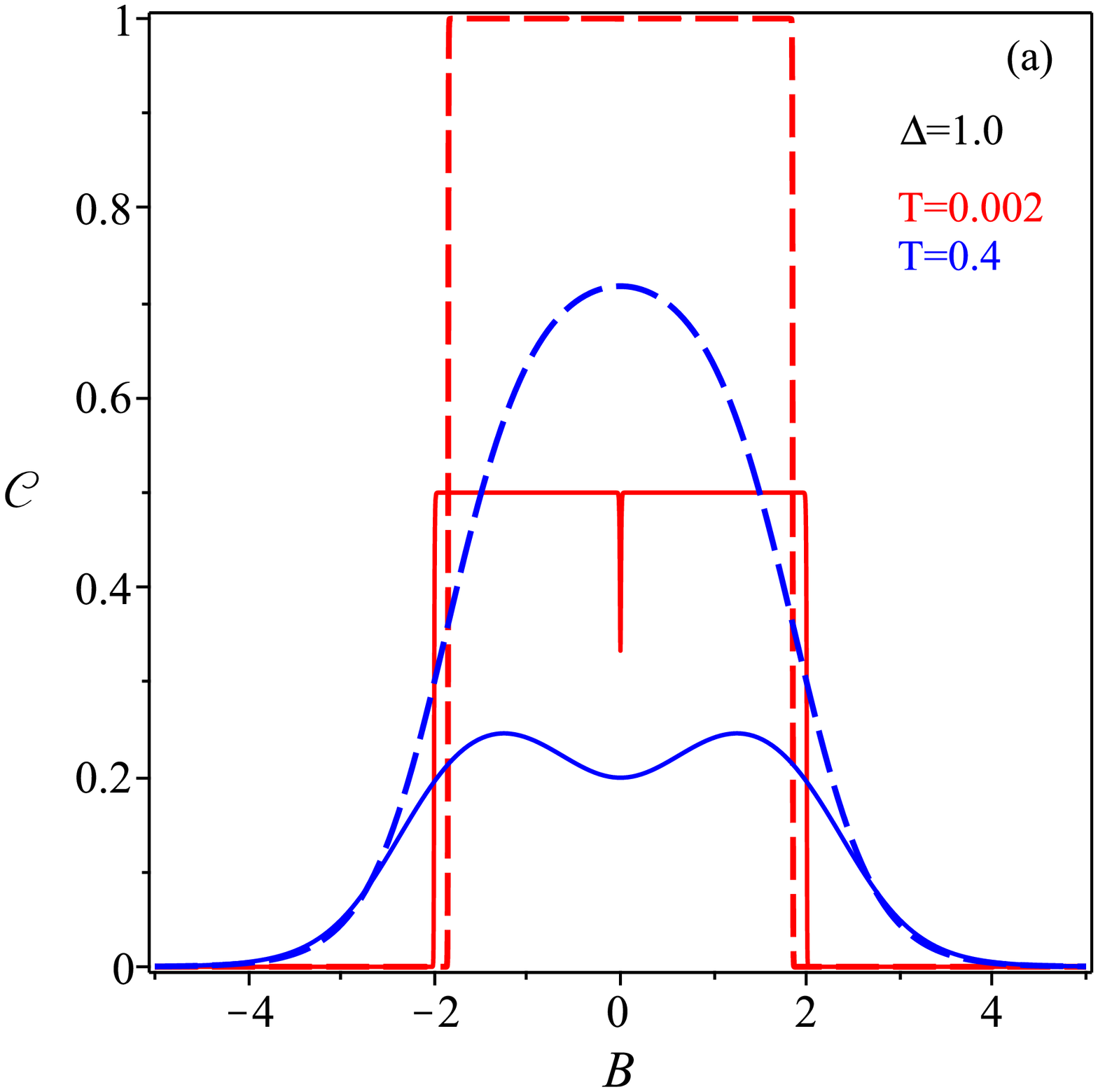}\includegraphics[scale=0.30]{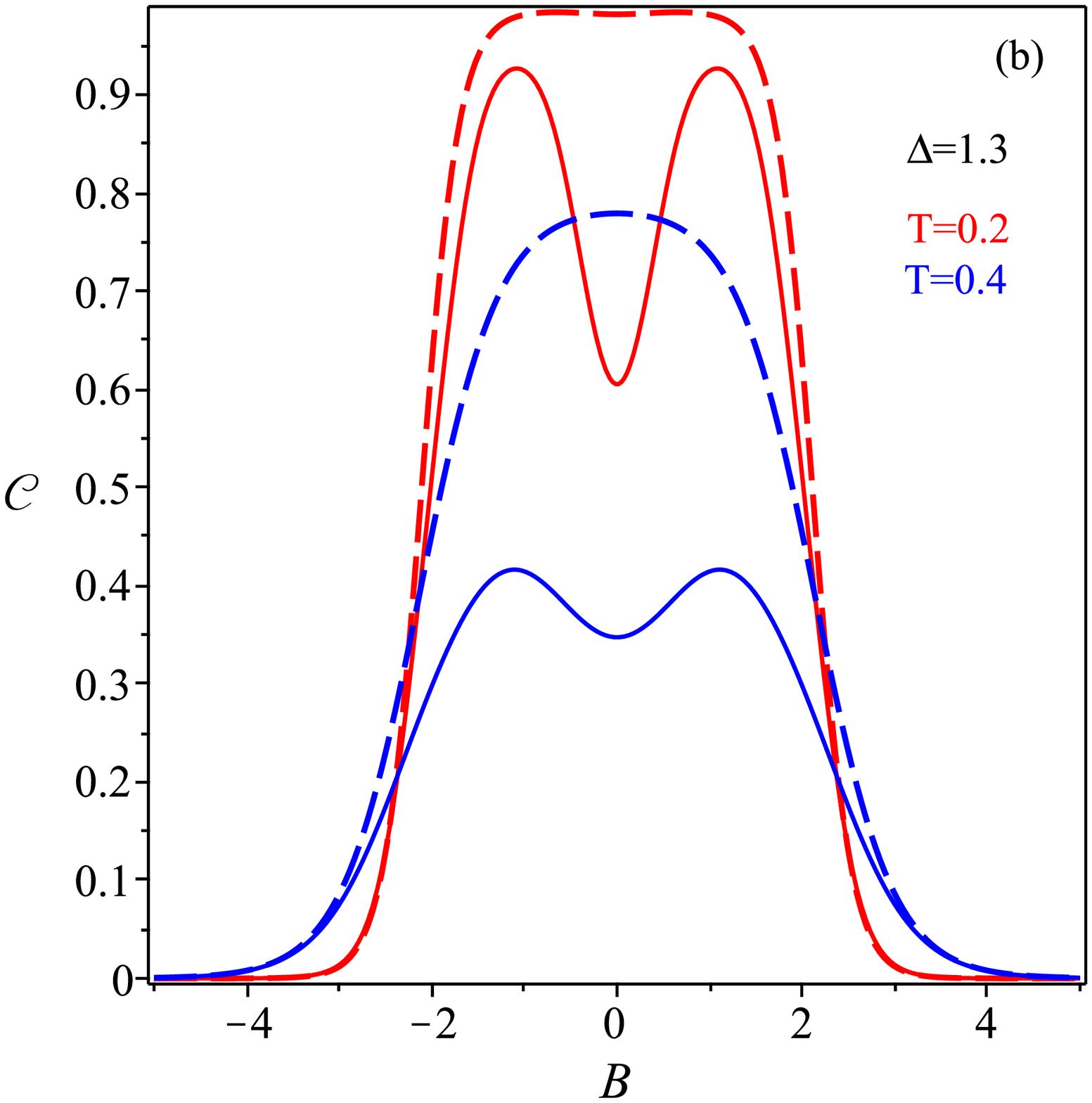}
	
	\caption{\label{fig6.1}The concurrence $\mathcal{C}$ as a function of the
		magnetic field $B$, we set $J_{0}/J=1.0$, $\alpha=0.0$, $\eta=-0.5$,
		$\gamma=-0.6$, $\Omega=0.8$. (a) dashed curve ($\Delta=1.0$, $T=0.002$,
		$T=0.400$), (b) dashed curve ($\Delta=1.3$, $T=0.200$, $T=0.400$)}
\end{figure}

Our results also show that with an appropriate choice of parameters our
model presents more robust entanglement for application of magnetic
field than the original model. In figure \ref{fig6.1} we present
some curves of the concurrence as a function of the magnetic field.
As before, here solid curves refer to the original model and the dashed
curves refer to the current model. Although the critical magnetic
field (above which the entanglement is zero) is the same for both
models, the entanglement is significantly greater in the current model.

\subsection{$l_{1}$-norm}
\begin{figure}\centering
	\includegraphics[scale=0.38]{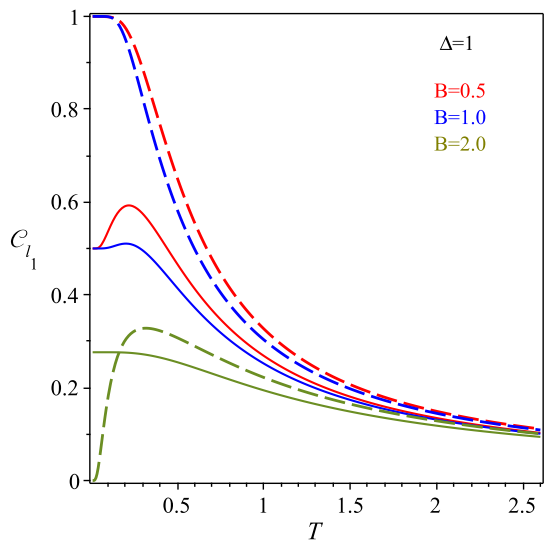}
	
	\caption{\label{fig:7}The quantum coherence $\mathcal{C}_{l_{1}}$ as a function
		of temperature $T$ for $J_{0}/J=1.0$, $\Delta=1.0$, $\alpha=0.0$,
		solid curve (original model), dashed curve ($\eta=-0.5$, $\gamma=-0.6$,
		$\Omega=0.8$).}
\end{figure}

Here we will discuss the results regarding the calculation of quantum
coherence using the $l_{1}$-norm. In Figure \ref{fig:7} we adopt
the same parameters adopted in Figure \ref{fig:2}. The comparison
between the two Figures is instructive. The behavior of the curves
is similar in the two Figures. Again there is a clear advantage of
the model with impurity (dashed curve) compared to the original model(solid
curve). The main difference is in the scale of magnitude. Quantum
coherence $\mathcal{C}_{l_{1}}$ is significantly greater than concurrence.
This result is not surprising since coherence measures, in addition
to entanglement, other quantum correlations.  Coherence is a 
basis-dependent measure and states without any quantum correlations can 
be fully coherent on an orthogonal basis. In addition, coherence is 
present also in single pure quantum systems which does not share any 
correlation with other systems.

\begin{figure}\centering
	\includegraphics[scale=0.38]{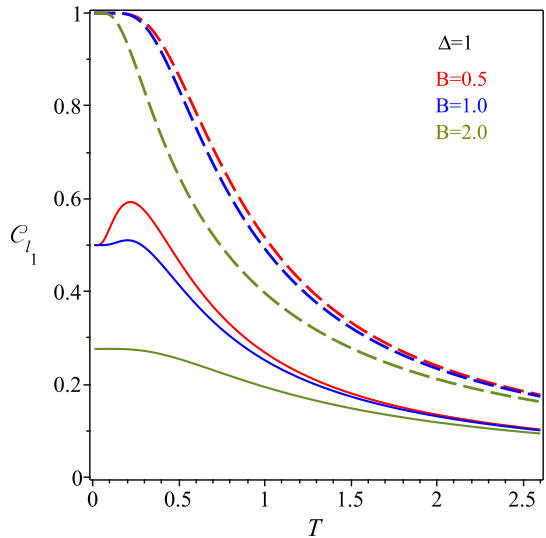}\caption{\label{fig:8}The quantum coherence $\mathcal{C}_{l_{1}}$ as a function
		of temperature $T$ for $J_{0}/J=1.0$, $\Delta=1.0$, solid curve
		(original model), dashed curve ($\alpha=0.5$, $\eta=-0.5$, $\gamma=-0.6$,
		$\Omega=0.8$).}
\end{figure}

Analogous to what was done in Figure \ref{fig:3}, in Figure \ref{fig:8}
we analyze how a change in the interaction between the spins of the
impurity affects the coherence $\mathcal{C}_{l_{1}}$. As with concurrence,
the increase in coherence with this small change was very significant.
The coherence of the impurity is greater than $0.2$ up to the temperature
of $2.5$ for all fields presented. The change in $\widetilde{J}$
provides considerable gain in both concurrence and coherence.

\begin{figure}
	\includegraphics[scale=0.30]{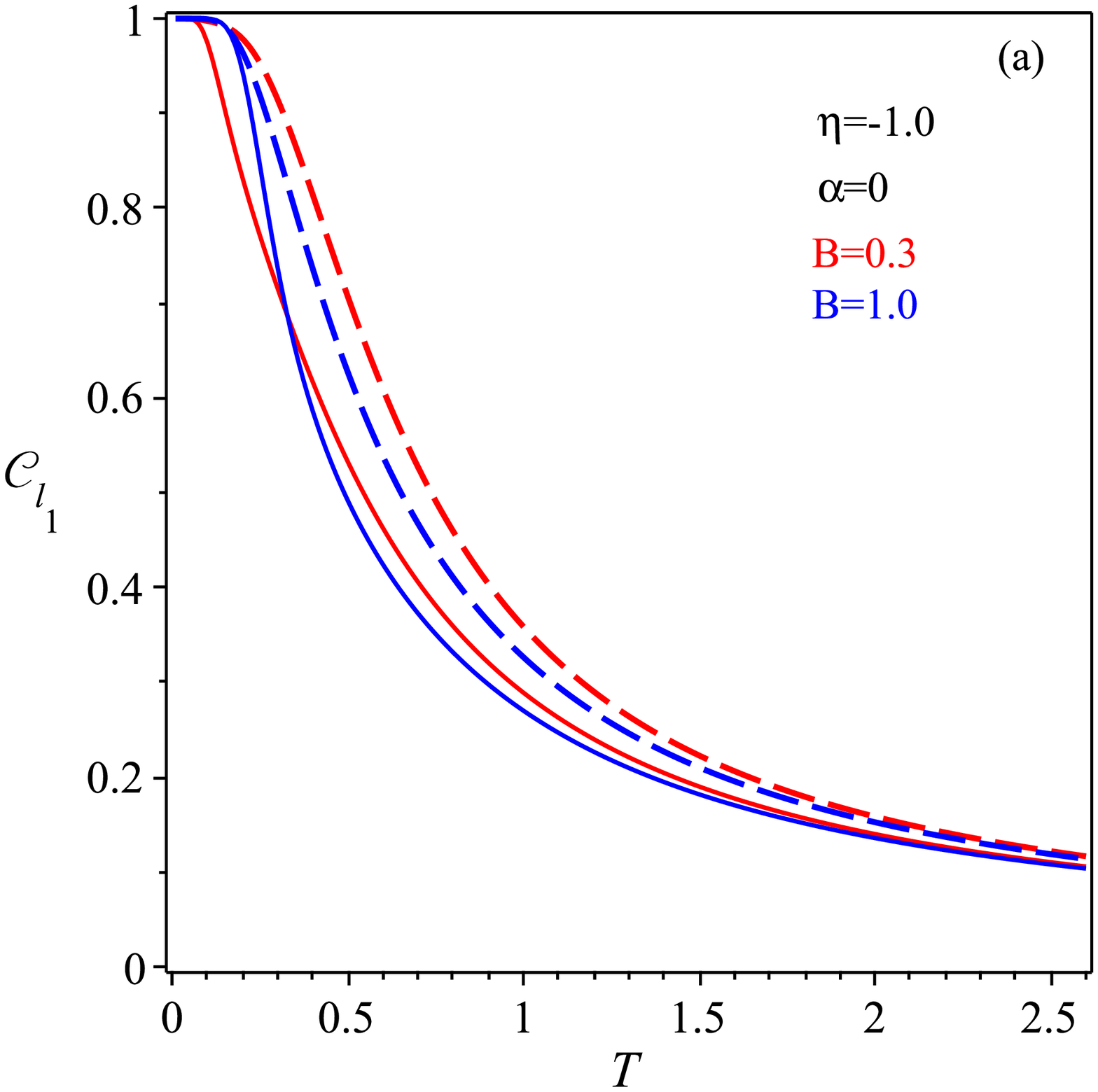}\includegraphics[scale=0.30]{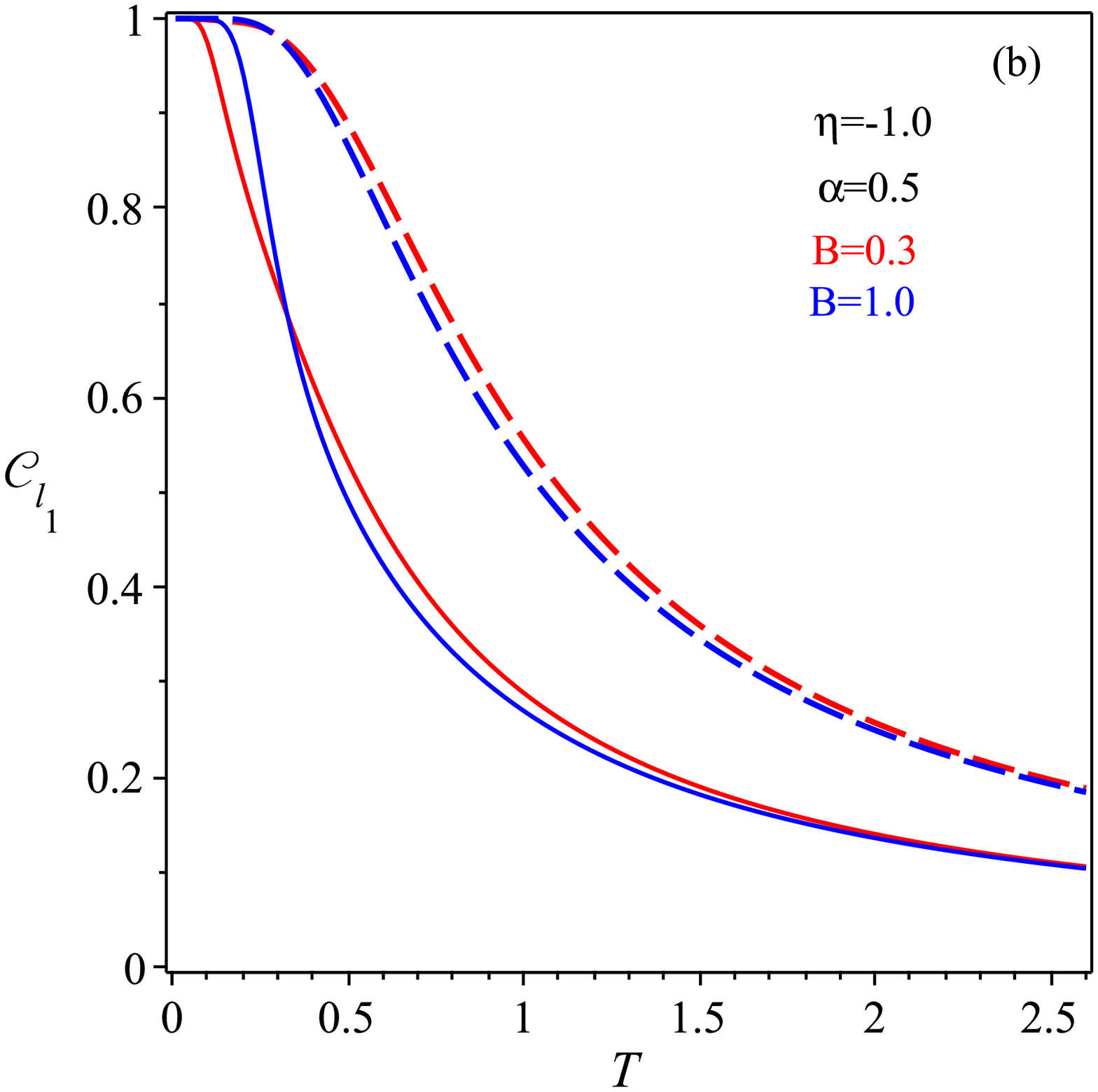}
	
	\caption{\label{fig:9}The concurrence $\mathcal{C}_{l_{1}}$ as a function
		of the temperature $T$, we set $J_{0}/J=1.0$, $\Delta=1.3$, $\eta=-1.0$,
		(a) dashed curve ($\alpha=0.0$, $\gamma=-0.8$, $\Omega=0.8$), (b)
		dashed curve ($\alpha=0.5$, $\gamma=-0.8$, $\Omega=0.8$)}
\end{figure}

In Figure \ref{fig:9} we will analyze the coherence by taking the
parameter $J_{1}=0.0$. It is interesting to note that in this case
the coherence of the original model and the model with impurity are
close when we take the parameter $\alpha=0.0$ (Figure \ref{fig:9}(a)).
From Figure \ref{fig5.1} it is evident that the same does not occur
with concurrence. This allows us to conclude that in this case the
coherence is proportionally greater than the concurrence in the original
model than in the model with impurity. In Figure \ref{fig:9}(b) we
can see the gain in coherence in relation to the original model increasing
the interaction between Heisenberg spins in impurity ($\alpha=0.5$).
\section{Quantum teleportation}

In this section, we study the quantum teleportation via a distorted
Ising-$XXZ$ diamond chain using the standard teleportation protocol.
By means of an entangled mixed state as resource, the standard teleportation
can be regarded as a general depolarizing channel \cite{peres} with
probabilities given by the maximally entangled components of the resource.
An unknown two-qubit pure state to be teleported can be written as $|\psi_{in}\rangle=\cos\left(\frac{\theta}{2}\right)|10\rangle+e^{i\phi}\sin\left(\frac{\theta}{2}\right)|01\rangle\;$, where $0\leq\theta\leq\pi$ and $0\leq\phi\leq2\pi$.
In the density operator formalism, the concurrence $\mathcal{C}_{in}$
of the input state, can be written as

\[
\mathcal{C}_{in}=2|e^{i\phi}\sin\left(\frac{\theta}{2}\right)\cos\left(\frac{\theta}{2}\right)|=|\sin(\theta)|\;.
\]

When a two-qubit state $\rho_{in}$ is teleported via the mixed channel
$\widetilde{\rho}_{ch}$, then the output state $\widetilde{\rho}_{out}$
is given by \cite{peres} 
\[
\widetilde{\rho}_{out}=\sum_{i,j=\left\{ 0,x,y,z\right\} }p_{i}p_{j}\left(\sigma_{i}\otimes\sigma_{j}\right)\rho_{in}\left(\sigma_{i}\otimes\sigma_{j}\right)\;,
\]
in which $p_{i}=tr\left[E^{i}\widetilde{\rho}_{ch}\right]$, $E^{0}=|\Psi^{-}\rangle\langle\Psi^{-}|$,
$E^{1}=|\Phi^{-}\rangle\langle\Phi^{-}|$, $E^{2}=|\Phi^{+}\rangle\langle\Phi^{+}|$
and $E^{3}=|\Psi^{+}\rangle\langle\Psi^{+}|$, where $|\Phi^{\pm}\rangle=\frac{1}{\sqrt{2}}\left(|00\rangle\pm|11\rangle\right)$
and $|\Psi^{\pm}\rangle=\frac{1}{\sqrt{2}}\left(|01\rangle\pm|10\rangle\right)$
are Bell states. Here, we consider the density operator channel as
$\widetilde{\rho}_{ch}\equiv\widetilde{\rho}(T)$. Then the output
density operator $\widetilde{\rho}_{out}$ is given by 
\begin{equation}
\widetilde{\rho}_{out}=\left[\begin{array}{cccc}
c & 0 & 0 & 0\\
0 & f & \chi & 0\\
0 & \chi & g & 0\\
0 & 0 & 0 & c
\end{array}\right]\;.\label{eq:rho-out}
\end{equation}
The elements of the operators can be expressed as 
\begin{flushleft}
	\[
	\begin{array}{cl}
	c= & \left(\widetilde{\rho}_{2,2}+\widetilde{\rho}_{3,3}\right)\left(\widetilde{\rho}_{1,1}+\widetilde{\rho}_{4,4}\right),\\
	f= & \left(\widetilde{\rho}_{1,1}+\widetilde{\rho}_{4,4}\right)^{2}\cos^{2}\left(\frac{\theta}{2}\right)+\left(\widetilde{\rho}_{2,2}+\widetilde{\rho}_{3,3}\right)^{2}\sin^{2}\left(\frac{\theta}{2}\right),\\
	g= & \left(\widetilde{\rho}_{2,2}+\widetilde{\rho}_{3,3}\right)^{2}\cos^{2}\left(\frac{\theta}{2}\right)+\left(\widetilde{\rho}_{1,1}+\widetilde{\rho}_{4,4}\right)^{2}\sin^{2}\left(\frac{\theta}{2}\right),\\
	\chi= & 2e^{i\phi}\widetilde{\rho}_{2,3}^{\,2}\sin\theta.
	\end{array}
	\]
	\par\end{flushleft}

According to the definition, the concurrence of the input state is
given as $\mathcal{C}_{in}=\sin(\frac{\theta}{2})$, and the output
one follows by 
\[
\mathcal{C}_{out}(\widetilde{\rho})=2max\left\{ 2\widetilde{\rho}_{2,3}^{\,2}\mathcal{C}_{in}-2|\widetilde{\rho}_{2,2}||\widetilde{\rho}_{1,1}-\widetilde{\rho}_{4,4}|,0\right\} .
\]

More recently, the teleportation of the same entangled state was studies
in this model without impurities \cite{tele} and with impurities
\cite{mro}.

\section{Average fidelity of teleportation}

\begin{figure}\centering
	\includegraphics[scale=0.38]{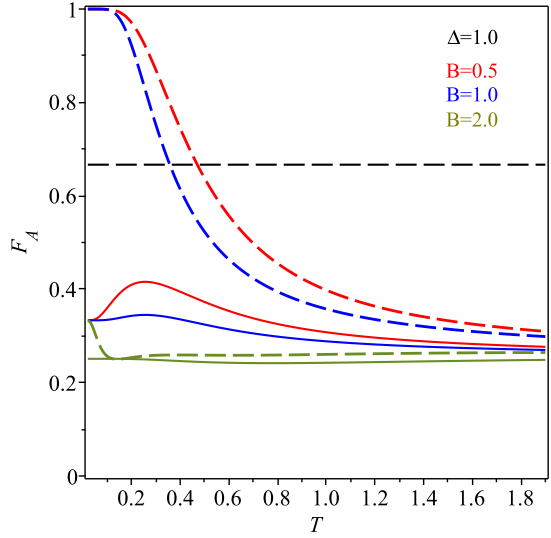}
	
	\caption{\label{fig:10}The average fidelity $F_{A}$ as a function of temperature
		$T$ for $J_{0}/J=1.0$, $\Delta=1.0$, solid curve (original model),
		dashed curve ($\alpha=0.0$, $\eta=-0.5$, $\gamma=-0.6$, $\Omega=0.8$).}
\end{figure}

In the following discussion, we turn our attention to the quality
of the entanglement teleportation. The quality of the teleportation
is measurement by the fidelity between the input state $\rho_{in}$
and the output state $\rho_{out}$. If the input state is pure, the
fidelity can be written as 
\[
F=\langle\psi_{in}|\rho_{out}|\psi_{in}\rangle\;.
\]
For this model $F$ can be written as 
\[
\begin{array}{cl}
F= & \frac{\sin^{2}\theta}{2}\left[\left(\widetilde{\rho}_{1,1}+\widetilde{\rho}_{4,4}\right)^{2}+4\widetilde{\rho}_{2,3}^{\,2}-\left(\widetilde{\rho}_{2,2}+\widetilde{\rho}_{3,3}\right)^{\,2}\right]+
\left(\widetilde{\rho}_{2,2}+\widetilde{\rho}_{3,3}\right)^{\,2}.\label{eq:fidelityaver}
\end{array}
\]

In general the state to be teleported is unknown, it is more useful
to calculate the average fidelity. Thus the average fidelity of teleportation
can be formulated as 
\[
F_{A}=\frac{1}{4\pi}\intop_{0}^{2\pi}d\phi\intop_{0}^{\pi}F\sin\theta d\theta\;.
\]
The average fidelity of teleportation can be written as 

\[
\begin{array}{cl}
F_{A}= & \frac{1}{3}\left[\left(\widetilde{\rho}_{1,1}+\widetilde{\rho}_{4,4}\right)^{2}+4\widetilde{\rho}_{2,3}^{\,2}-\left(\widetilde{\rho}_{2,2}+\widetilde{\rho}_{3,3}\right)^{\,2}\right]+
\left(\widetilde{\rho}_{2,2}+\widetilde{\rho}_{3,3}\right)^{\,2}.\label{eq:fidelityaverage}
\end{array}
\]

\begin{figure}\centering
	\includegraphics[scale=0.38]{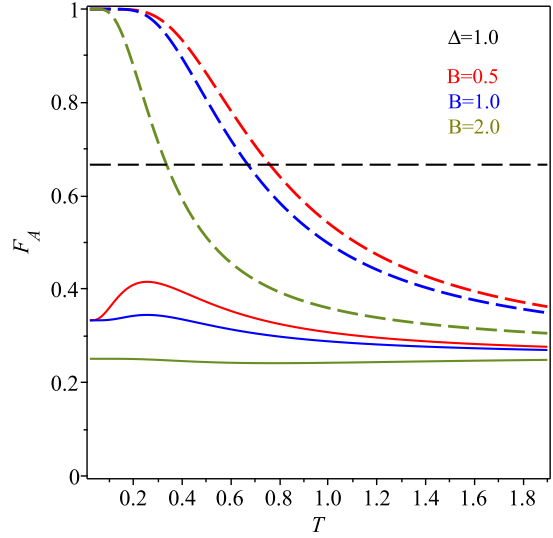}\caption{\label{fig:11} The average fidelity ${F}_{A}$ as a function of temperature
		$T$ for $J_{0}/J=1.0$, $\Delta=1.0$, solid curve (original model),
		dashed curve ($\alpha=0.5$, $\eta=-0.5$, $\gamma=-0.6$, $\Omega=0.8$).}
\end{figure}

To transmit a quantum state $|\psi_{in}\rangle\;$ better than any
classical communication protocol, $F_{A}$ must be great than $\frac{2}{3}$
which is the best fidelity in the classical world \cite{joz}. To
demonstrate the effects of the impurity on the average fidelity, we
plot the behavior of average fidelity. In Fig. \ref{fig:10},  we show
the average teleportation $F_{A}$ as a function of temperature $T$
for $J_{0}/J=1.0$, $\Delta=1.0$ and several values of magnetic field.
For the model with impurity, we set $\alpha=0.5$, $\eta=-0.5$, $\gamma=-0.6$,
$\Omega=0.8$. The horizontal dashed lines at $F_{A}=\frac{2}{3}$
denote the limit of quantum fidelities. We can see that in the original
model the average fidelity is below $\frac{2}{3}$, signaling that
it is not possible to teleport information. Meanwhile, when consider
the impurity, we have a considerable improvement in quantum teleportation.
For weak magnetic field, the average fidelity reaches maximum value
at low temperatures as the temperature increases, the average fidelity
decays up to the critical temperature, beyond which the teleportation
of information is no longer valid. On the other hand, for strong magnetic
field, the effect of the impurity on the quantum teleportation does
not occurs. It means that, average fidelity remains below $\frac{2}{3}$
regardless of the temperature rise.

In Fig. \ref{fig:11}, the behavior of the average fidelity as a function
of temperature $T$ with fixed parameters values, $J_{0}/J=1.0$,
$\Delta=1.0$ $\eta=-0.5$, $\gamma=-0.6$, $\Omega=0.8$ and different
values of magnetic field $B$ is depicted. In this figure, we also
consider the impurity in the parameter $\alpha$ of Heisenberg exchange
interaction, we select $\alpha=0.5$, under this condition a considerable
increase in average fidelity is observed, reaching its maximum value
$F_{A}=1$ for several values of magnetic field, including strong
magnetic fields (dashed curves). After that, it gradually decreases
with increase the temperature. However, the average fidelity is always
less than $F_{A}<\frac{2}{3}$ for original model (solid curves),
that is, in this case the quantum teleportation fails. All these show
again that considerable enhancement of teleportation can be achieved
by tuning the strenght of impurity parameters.

\begin{figure*}
	\includegraphics[scale=0.30]{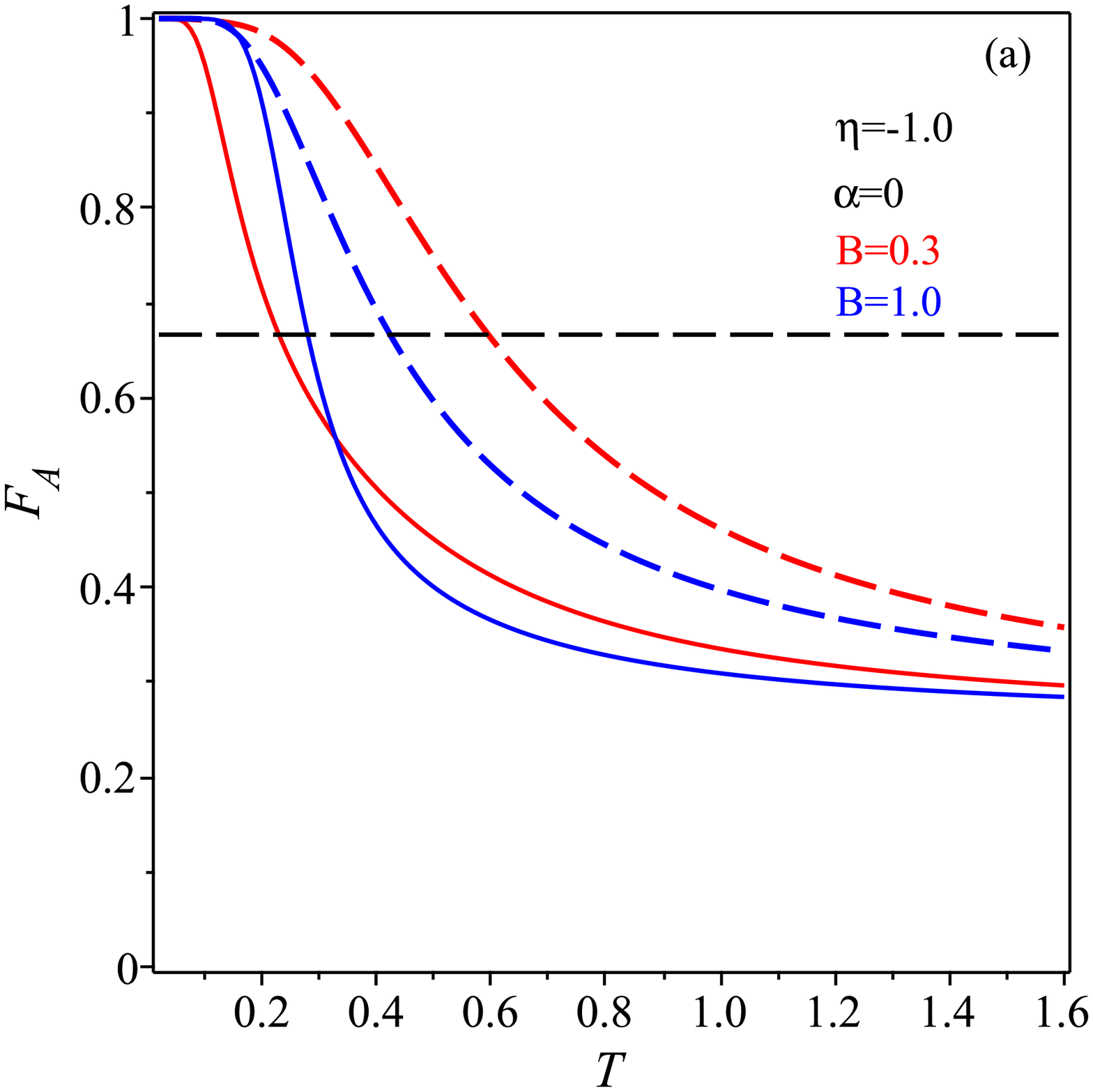}\includegraphics[scale=0.30]{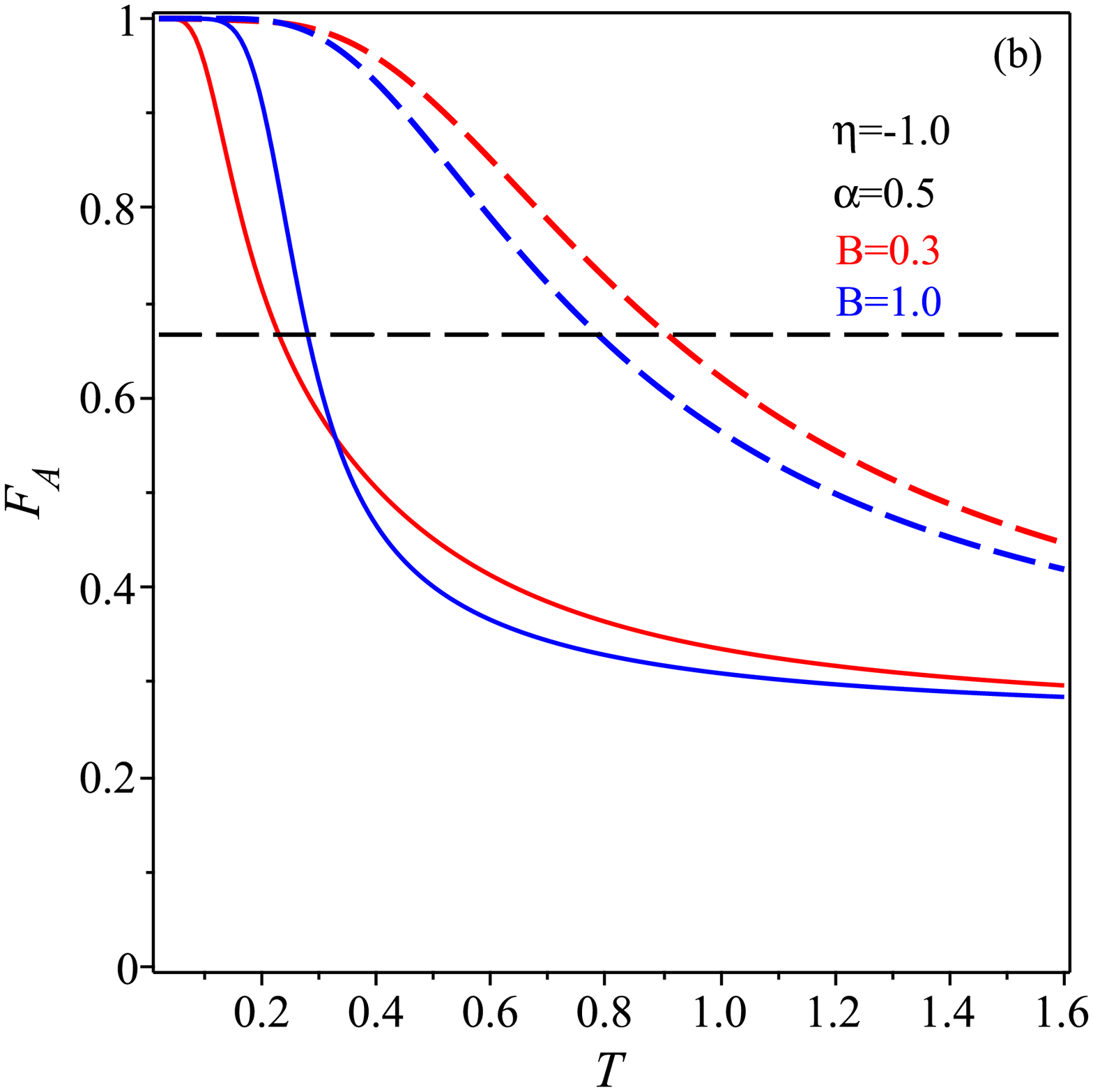}
	
	\caption{\label{fig:12}The average fidelity $F_{A}$ as a function of the
		temperature $T$, we set $J_{0}/J=1.0$, $\Delta=1.3$, $\eta=-1.0$,
		(a) dashed curve ($\alpha=0.0$, $\gamma=-0.8$, $\Omega=0.8$), (b)
		dashed curve ($\alpha=0.5$, $\gamma=-0.8$, $\Omega=0.8$)}
\end{figure*}

Finally, in Fig. \ref{fig:12}, let us study the behavior of the system
in extreme cases. We analyze the case when $J_{1}=0.0$ or $\eta=-1.0$
which corresponds to collapse of one of the impurities of the Ising-like
interaction, as depicted in Fig. \ref{fig:4}. We set $J_{0}/J=1.0$,
$\Delta=1.3$. In Fig. \ref{fig:12}(a), we illustrate the average
fidelity as a function of temperature $T$ for different values of
the magnetic field and the fixed values of impurities parameters $\alpha=0.0$,
$\gamma=-0.8$, $\Omega=0.8$. As can be seen, the results of $F_{A}$
indicate that, for low temperatures, the average fidelity is maximum,
that is, $F_{A}=1$, for both models. However, for the higher temperature,
average fidelity with impurity is more robust in comparison to that
without it.

On the other hand, Fig. \ref{fig:12}(b) shows the average fidelity
for two values of magnetic field and we fixed $\alpha=0.5$, $\gamma=-0.8$,
$\Omega=0.8$. This figure shows clearly that the behavior the average
fidelity is even more robust when we consider the impurity in the
Heisenberg spin exchange parameter $\alpha=0.5$, allowing the teleportation
of information at higher temperature. Altogether, it could be concluded
that impurity generated by a local bound disorder of the interchain
exchange coupling affects strongly the quantum correlations, which
generates a considerable increase in the teleportation of information.
\section{Conclusions}

In summary, we have investigated the effects due to the inclusion
of an distorted impurity plaquette on the spin-1/2 Ising-$XXZ$ diamond
chain. Next, the quantum correlations and teleportation of the model
under investigation have been exactly calculated using the transfer-matrix
technique. In particular, our attention has been paid to a rigorous
analysis of the quantum correlations of the interstitial distorted
impurity Heisenberg dimers through the concurrence and norm-$l_{1}$.
We observed that the impurity strongly influence at the behavior of
the entanglement and quantum coherence. More specifically, the results
show the existence of the thermal entanglement and quantum thermal
coherence in regions beyond the reach of the original model. One of
our notable results is that quantum correlations can be controlled
and tuned by Ising and Heisenberg parameters into model.

In addition, we have also discussed the teleportation of the two-qubits
in an arbitrary state through a quantum channel composed by a couple
of Heisenberg dimers with distorted impurity in an Ising-$XXZ$ diamond
chain structure. We have found that, the quantum teleportation increases
when increases some suitable impurity parameters. The influence of
impurity is more evident in the average fidelity when we consider
the impurity in the Heisenberg parameter.

Finally, our results show that the impurities can be manipulated to
locally control the thermal quantum correlations, unlike the original
model where it is done globally.

\section*{Acknowledgments}
This work was partially supported by CNPq, CAPES and Fapemig. M. Rojas
would like to thank CNPq grant 432878/2018-1.

%
%

\bibliographystyle{elsarticle-num}

\end{document}